\documentclass{vgtc}                          

\ifpdf
  \pdfoutput=1\relax                   
  \usepackage{graphicx}                
  \DeclareGraphicsExtensions{.pdf,.png,.jpg,.jpeg} 
\else
  \usepackage{graphicx}                
  \DeclareGraphicsExtensions{.eps}     
\fi%

\graphicspath{{figures/}{pictures/}{images/}{./}} 
\usepackage{multirow}
\usepackage{amssymb}
\usepackage{xspace}
\usepackage{microtype}                 
\PassOptionsToPackage{warn}{textcomp}  
\usepackage{textcomp}                  
\usepackage{mathptmx}                  
\usepackage{times}                     
\usepackage{cite}                      
\usepackage{tabu}                      
\usepackage{booktabs}                  
\usepackage{hyperref}
\usepackage{cleveref}
\usepackage{amssymb}
\usepackage[normalem]{ulem}

\usepackage[dvipsnames]{xcolor}
\usepackage{diagbox}
\definecolor{Myyellow}{HTML}{E2BE04}
\definecolor{Mygrey}{HTML}{666666}
\definecolor{gray}{HTML}{666666}
\definecolor{Myorange}{HTML}{ef6c00}
\definecolor{Myblue}{HTML}{4a86e8}
\definecolor{Mygreen}{HTML}{60B177}
\definecolor{Myyellow2}{HTML}{fffeb2}
\definecolor{Mypurple}{HTML}{ba6d9a}

\newcommand{\gray}[1]{\textcolor{gray}{#1}}
\newcommand{\green}[1]{\textcolor{black}{#1}}

\newcommand{\stitle}[1]{\noindent\textbf{#1}\xspace}
\newcommand{\edit}[1]{\textcolor{black}{#1}}

\newenvironment{myitemize}{\begin{list}{\gray{$\bullet$}}{\renewcommand{\leftmargin}{0.15in}}}{\end{list}}

\onlineid{0}

\vgtccategory{Research}

\vgtcinsertpkg

\title{How Do Captions Affect Visualization Reading?}

\author{Hanxiu `Hazel' Zhu \thanks{These authors contributed equally.}~  {\thanks{e-mail: \{hazel.zhu, shelly.cheng\}@columbia.edu}}   %
\and Shelly Shiying Cheng \footnotemark[1] ~\footnotemark[2]
\and Eugene Wu \thanks{e-mail: ewu@cs.columbia.edu}}
\affiliation{\scriptsize Columbia University}

\abstract{Captions help readers better understand visualizations. However, if the visualization is intended to communicate specific features, should the caption be statistical, and focus on specific values, or perceptual, and focus on general patterns? Prior work has shown that when captions mention visually salient features, readers tend to recall those features. Still, we lack explicit guidelines for how to compose the appropriate caption.  Further, what if the author wishes to emphasize a less salient feature?  

In this paper, we study how the visual salience of the feature described in a caption, and the semantic level of the caption description, affect a reader’s takeaways from line charts. For each single- or multi-line chart, we generate 4 captions that 1) describe either the primary or secondary salient feature in a chart, and 2) describe the feature either at the statistical or perceptual levels. We then show participants random chart-caption pairs and record their takeaways.  

We find that the primary salient feature is more memorable for single-line charts when the caption is expressed at the statistical level; for primary and secondary features in multi-line charts, the perceptual level is more memorable. We also find that many readers will tend to recall y-axis numerical values when a caption is present.}

\CCScatlist{
  \CCScatTwelve{Human-centered computing}{Visu\-al\-iza\-tion}{Visualization design and evaluation methods}{}
}

\begin{document}


\firstsection{Introduction}

\maketitle


Visualizations are used to communicate important data trends and patterns, but can be hard to understand and remember. Captions help the reader better understand the visualization, but how should they be written? And how do they affect the reader's takeaways?

Recent work studied how a reader's takeaways after reading a visualization are affected by the visual salience of the data described in the caption~\cite{Kim2021TowardsUH} and the semantic content in the caption~\cite{2022-vis-text-model}.   Specifically, Lundgard et al.~\cite{2022-vis-text-model} find that such content can be classified into four semantic levels: 1) descriptions of the visualization encoding; 2) statistical descriptions of values encoded in the visualization; 3) perceptual descriptions of trends, patterns, or outliers; and 4) contextual knowledge not present in the visualization.  

Of the four semantic levels, {\it statistical} and {\it perceptual} are most relevant when composing captions that {\it describe data encoded in the visualization}.  \edit{However, in many cases, the visualization author wants to highlight a particular feature in the visualization for the reader. In these conditions, what semantic level should be used, and do factors such as the visual salience of the feature or the type of chart play a role?}


\edit{This paper reports an initial study towards answering this question.}
We present participants with single- and multi-line charts; each chart's caption describes either the \edit{most visually salient feature (primary feature) or a non-primary feature (secondary)}, and is written at either the statistical or the perceptual level.  We then study how the participant's takeaways relate to the content and emphasis in the chart caption. \edit{Specifically, we hypothesize that the semantic level and the feature described in a caption can affect if a takeaway mentions the primary or secondary feature of single- or multi-line charts.}

We find that for single-line charts, the primary features are already prominent and memorable, and yet participants are even more likely to recall it when a statistical level caption is present.  The secondary features of single-line charts are still relatively salient, and both statistical and perceptual level captions can make the features more memorable to the participants.  In contrast, multi-line charts are typically crowded, so both primary and secondary features are best described at the perceptual level. \edit{Regardless of chart types or feature salience, captions help participants memorize the features they describe. }

Based on our findings, we provide guidelines to help data visualization practitioners write better captions to communicate the feature they want to highlight. Our guidelines can also be used to help standardize machine-generated captions.





\section{Related Work}\label{s:related}

Various accessible media publishers have established general guidelines for people to write natural language captions \cite{diagramcenter, GBH, CFPB, W3C}, but those guidelines often lack rationales and support from empirical experiments~\cite{Jung2021CommunicatingVW}. This critique motivates our empirical study to better understand how to write effective data captions.


\subsection{The Impact of Visual Salience on Reader Takeaway}
Prior work has shown that the salience of the feature described by a caption affects what the reader takes away. Kim et al.~\cite{Kim2021TowardsUH} show that, for single-line charts, when the caption mentions a salient feature, reader takeaways more consistently mention the feature; when the caption mentions a less salient feature, reader takeaways are more likely to mention the most salient feature than the feature described in the caption. In our study, we investigate how visual salience and semantic level of a caption affect the reader's takeaway. We also offer recommendations when practitioners want to emphasize less salient features. 

\subsection{Natural Language Models for Captions}
The lack of rationale for caption guidelines raises the need to analyze captions systematically. While the three-level model by Kim et al.~\cite{Kim2021AccessibleVD} guides people on how to scaffold visualization information in order,  Lundgard et al.~\cite{2022-vis-text-model} propose a concrete model that categorizes the content in a caption into four semantic levels: 1) elemental and encoded, 2) statistical and relational, 3) perceptual and cognitive, and 4) contextual and domain-specific (\Cref{t:lundgard}).

\begin{table}[h]
\includegraphics[width=8.5cm]{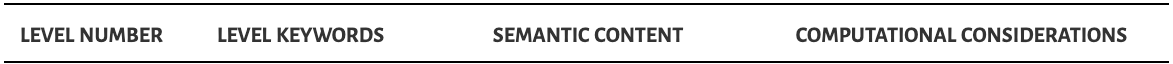}
\includegraphics[width=8.5cm]{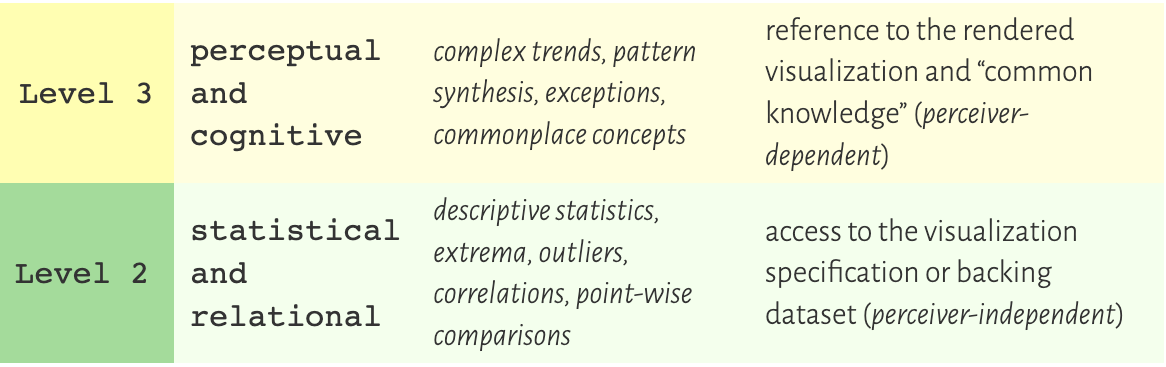}
\caption{Statistical and perceptual levels of Lundgard et al.'s~\cite{2022-vis-text-model} Model of Semantic Content.}
\label{t:lundgard}
\end{table}

Lundgard et al.~\cite{2022-vis-text-model} evaluate the effectiveness of each semantic level, and suggest that both sighted and visually impaired readers tend to regard semantic levels that communicate perceptual information (level 3) and statistical information (level 2) as useful to an extent. Although sighted users also ranked contextual level (level 4) highly, this level depends on the reader's subjective knowledge about the world events (context).   Our work teases out the nuances between captions that communicate data {\it in the chart}---in other words, the perceptual and statistical levels.



\subsection{Evaluating Captions}

Some methods of evaluating captions used in the past empirical research include semi-structured interviews\cite{Jung2021CommunicatingVW}, pre-constructed questions\cite{Moraes2014EvaluatingTA, Kildal2006NonvisualOO}, and takeaways recall\cite{Kong2019TrustAR, Borkin2016BeyondMV, Kim2021TowardsUH}. 

Semi-structured interviews allow users to express their thoughts and preferences directly after reading captions\cite{Jung2021CommunicatingVW}. While such methods can provide detailed and insightful qualitative results, it is challenging to generalize meaningful quantitative results at scale.

Moraes et al.\cite{Moraes2014EvaluatingTA} and Kildal et al.\cite{Kildal2006NonvisualOO} ask users to listen to captions or data values, and answer some pre-constructed overview questions after that. The time taken to answer those questions and the correct answers are used to evaluate the effectiveness of the caption or tool used. This method is helpful for quantitatively evaluating captions that provide a large picture of visualizations. However, it is not suited for captions categorized under Lundgard et al.’s model, which focus on specific visualization features.

Kong et al. \cite{Kong2019TrustAR}, Borkin et al.\cite{Borkin2016BeyondMV} and Kim et al.\cite{Kim2021TowardsUH} ask users to recall their takeaways from some visualizations or their captions, and analyze the takeaways. We find this method to be the most relevant to our study, and follow the takeaways recall method similar to Kim et al.'s \cite{Kim2021TowardsUH}. However, the purpose of our study differs largely from Kim et al., who focus on what information to put in captions. Our study focuses on how varying semantic levels of caption content can affect users' understanding of charts and captions.
\section{Methods}\label{s:methods}


We study how captions written at the statistical or perceptual levels affect the memorability of visualization features.  Since these semantic levels are based on Lundgard et al.'s model~\cite{2022-vis-text-model}, we use two single- and two multi-line charts from their corpus~\cite{mitdataset}. 

We follow the 3-step process used by Kim et al.~\cite{Kim2021TowardsUH}: given a visualization, we (Step 1) identify visually salient features, (Step 2) generate captions for the primary and secondary most salient features, and finally (Step 3) show captioned visualizations to participants and ask them to write their takeaways. The study was approved by our institution’s IRB.


\subsection{Step 1: Identify visually salient features}

\begin{figure}[h]
\includegraphics[width=8.5cm]{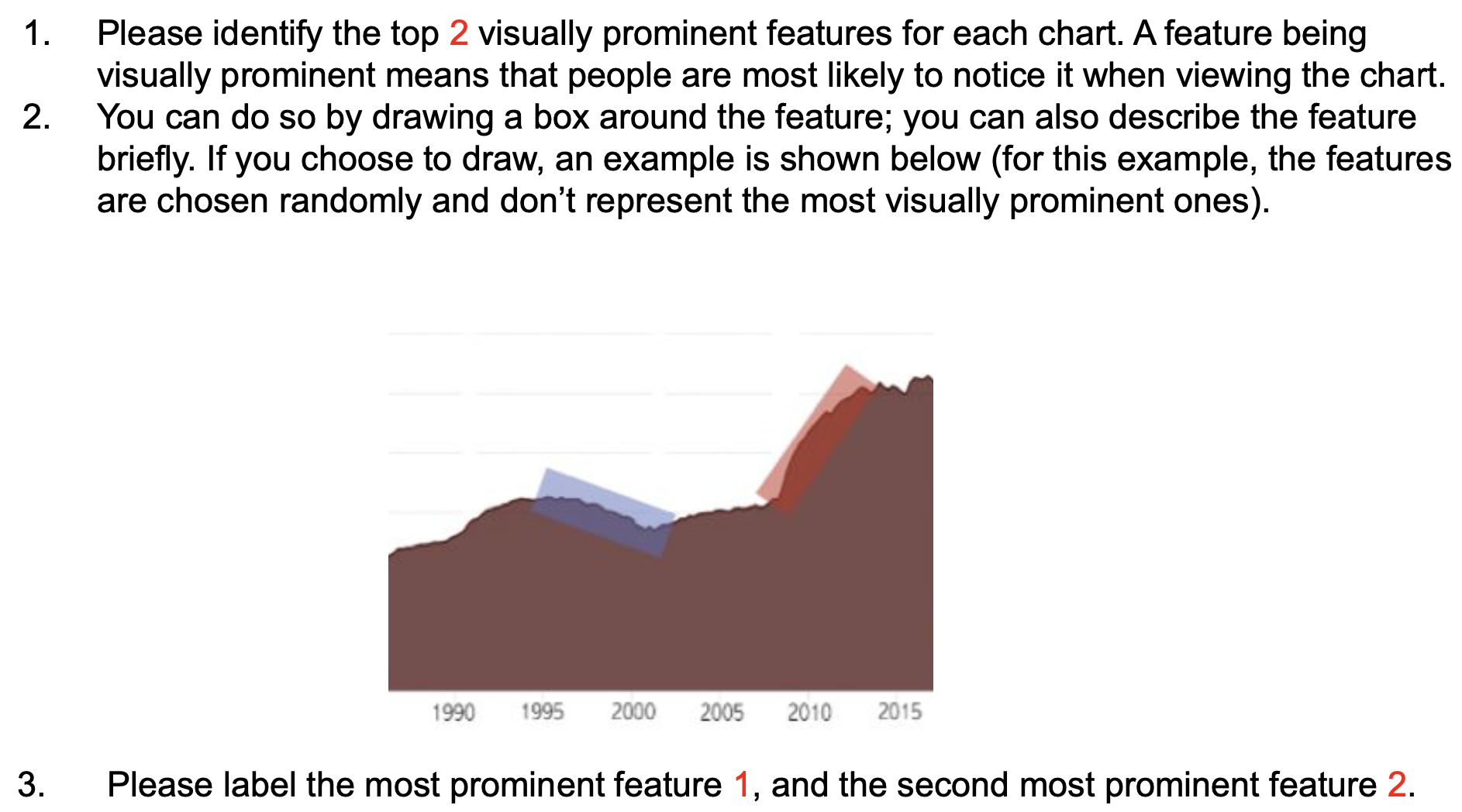}
\caption{Instructions for identifying visually salient features.}
\label{f:study1}
\end{figure}

We first determined the primary and secondary most salient visualization features using 7 charts from \edit{from Lundgard et al.'s corpus\cite{mitdataset}}.     We showed 9 university students each chart, and asked them to draw boxes and label each box as primary or secondary.  
They could also write a short description if they felt drawing a box is inadequate. The instructions for this step are shown in \Cref{f:study1}. We then aggregated their submissions to choose the primary and secondary visual features for participants in steps 2 and 3 of the study.

\begin{figure}[h]
\includegraphics[width=8.5cm]{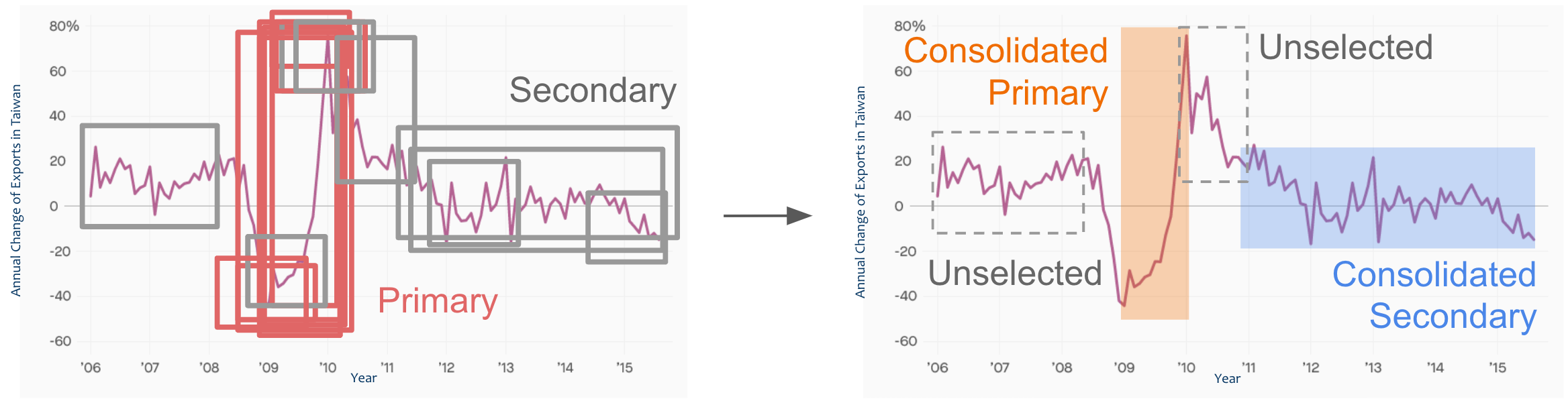}
\caption{An example of identified primary and secondary features from 9 participants' responses. We first cluster the bounding boxes, \edit{and choose the top two clusters as the primary and secondary features.   Here, the primary feature (\textcolor{Myorange}{orange}) describes the rise from minimum in 2009 to maximum in 2010; the secondary feature (\textcolor{Myblue}{blue}) covers the relatively stable trend after 2011. If the top clusters overlap, we discard the chart. }}
\label{f:boxes}
\end{figure}

\Cref{f:boxes} shows the primary and secondary boxes drawn for one of the charts.    We clustered \edit{boxes as follows: if a new box overlaps with a box in an existing cluster by 80\% or more, then we add it to the cluster, otherwise it starts a new cluster.} We then chose regions that overlapped with the top two clusters and used those regions as the primary (in \textcolor{Myorange}{orange}) and secondary (in \textcolor{Myblue}{blue}) features of the chart.  

\begin{figure*}[h]
\includegraphics[width=\textwidth]{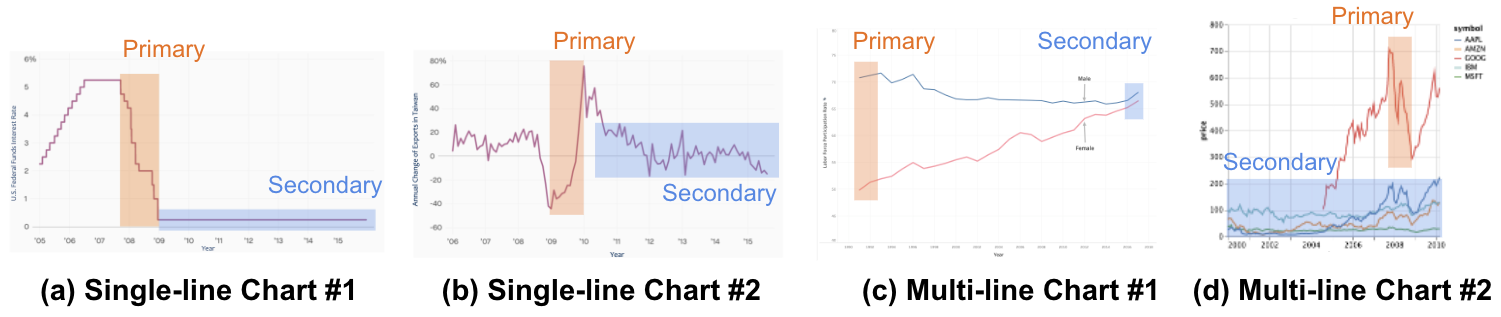}
\caption{Single- and multi-line charts with identified salient features (primary in \textcolor{Myorange}{orange} and secondary in \textcolor{Myblue}{blue} highlight) used in step 2 and step 3.}
\label{f:charts}
\end{figure*}

\edit{It was important that there was consensus about the primary feature, so we discarded a chart if the top two clusters of the chart are ambiguous.   It was less important whether there was considerable agreement about the secondary feature, as long as it was clearly distinguished from the primary feature.}
 This process resulted in 4 final charts: 2 single-line charts, 2 multi-line charts (\Cref{f:charts}). 
 
 \edit{To control for chart complexity, we referred to Lundgard et al.\cite{mitdataset}, who classified the 2 single-line charts in \Cref{f:charts} as {\it easy} in terms of complexity, and the 2 multi-line charts as {\it medium} complexity.   We considered more sophisticated ways to quantify complexity, such as Ryan et al.~\cite{ryan2018glance}, but they do not support multi-line charts, and so we leave this to future work. }
 
\subsection{Step 2: Generate Captions}
\edit{A major focus of this work is to evaluate the semantic levels proposed by Lundgard et al.~\cite{2022-vis-text-model}.   As such, 
we generated one caption at each semantic level for each feature in each chart.  A statistical caption is meant to describe 
the feature in the visualization using a numeric description, whereas a perceptual caption is meant to describe general trends.}   
With close reference to Lundgard et al.'s model~\cite{2022-vis-text-model}, we adhered to the following guidelines \edit{to create captions for the charts in \Cref{f:charts}.} 
%
\begin{myitemize}
 \item \stitle{Statistical:} Follow the general template: At [year], [dependent variable] is [value]. Insert ``lowest" or ``highest" in grammatically appropriate places if the feature contains an extremum.
 \item \stitle{Perceptual:} Use one or more keywords that describe trends and patterns: increase, decrease, drop, rebound, stable, trend, gap. Add values in grammatically appropriate places so that both levels are comparable in granularity.
\end{myitemize}
%
%
The independent variables are features described in the caption (primary or secondary) and the semantic level of the captions (statistical or perceptual), and the dependent variable is the user takeaway.
For each chart, we generated 4 captions (statistical or perceptual $\times$ primary or secondary feature), along with a no-caption control group.  As an example, the captions for \Cref{f:charts}(b) are:

\begin{myitemize}
    \item \textcolor{Myorange}{\bf Primary}, \textcolor{Mygreen}{\bf{Statistical}}: At 2009, Taiwan has the lowest change of exports of -45\%; at 2010, it has the highest change of exports of 75\%. 

    \item \textcolor{Myorange}{\bf Primary}, \textcolor{Myyellow}{\bf{Perceptual}}: After a dramatic drop in exports in 2009, the percentage of exports rebounds significantly and reaches the peak of 75\% in 2010.

    \item \textcolor{Myblue}{\bf Secondary}, \textcolor{Mygreen}{\bf{Statistical}}: At 2011, the annual change of exports in Taiwan is 30\%; at 2016, it is -18\%.
    \item \textcolor{Myblue}{\bf Secondary}, \textcolor{Myyellow}{\bf{Perceptual:}} After 2011, the exports become more stable, but the overall trend decreases, with exports reaching -18\% in 2016.
\end{myitemize}

\subsection{Step 3: Collect Takeaways for Charts \& Captions}
\begin{figure}[h]
\includegraphics[width=8.5cm]{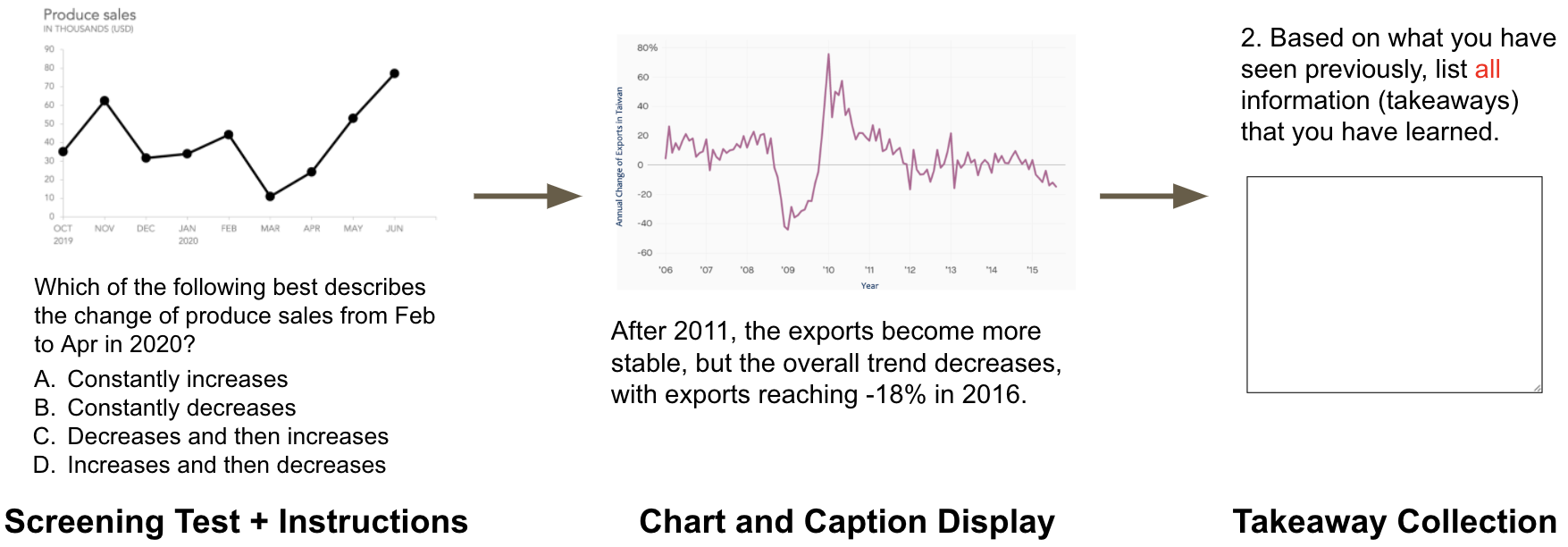}
\caption{Procedures for Collecting Takeaways.}
\label{f:procedures}
\end{figure}

Finally, we showed the captioned visualizations to a different set of participants and collect their takeaways.   We recruited participants from three main channels: a university campus, Amazon Mechanical Turk (AMT), and visualization-related online communities. Participants were asked to fill an online questionnaire (\Cref{f:procedures}). 

The questionnaire starts with a screening test to ensure that a participant can read values and trends from a visualization.  Each participant then reads 4 charts in random order, with a random caption version for each chart. \edit{For the final data collected, caption types (including the no caption control group) have a balanced distribution, with each type constituting roughly 20\% of all captions.} 


To mimic real-world settings, the participant can read the chart-caption pair as long as they wish, and then click ``next'' to write their takeaways. We don't allow participants to go back to the previous page to \edit{reduce the likelihood that they copy-and-paste the caption into their takeaways.} This condition is made clear for every chart.  We also ask the participants to report their reliance on the chart and caption when writing their takeaways, based on a 5-point Likert scale.


The first two authors of this paper coded the takeaways based on the criteria in \Cref{t:coding}.  Each author labeled the takeaways independently and discussed confusing cases together. 
\edit{If a participant copied captions, entered ``NA", or clearly showed that they misunderstood the questionnaire in any takeaway they wrote,  we disqualified all takeaways from the participant.} \edit{For example, one participant wrote ``It's important to know interest rate" for Figure 3(a), which was not something that can be obtained by looking at the chart and the caption. We also disqualified a participant if their takeaway was extremely vague. For example, one participant wrote ``increase and decrease" for all four charts, and we couldn't map the takeaway to a specific feature of the chart. Among the 124 completed responses that we received, we disqualified 23.}

\begin{table}[h]
\centering
\begin{tabular}{ll}
 \textbf{Label} & \textbf{Explanation} \\ 
   \midrule
 Primary & If takeaway mentions primary feature. \\ 
 Secondary & If takeaway mentions secondary feature.\\
 Other & If takeaway mentions any other feature. \\
 Numeric-x & If takeaway contains x-axis values. \\
 Numeric-y & If takeaway contains y-axis values. \\
\end{tabular}
\vspace{1mm}
\caption{Labels for possible information in takeaways.}
\label{t:coding}
\end{table}



\section{Analysis}\label{s:results}

 
In total, we collected 404 takeaways from 101 qualified participants. 92\% were 18-54 years old, 88\% held a Bachelor's degree or higher, and 92\% agreed that they read visualizations proficiently in everyday life.  Assuming that captions are more effective at communicating a given feature in the data when the participant mentions the feature in their takeaway, we quantify and report {\it memorability} as the percentage of takeaways that mentioned the feature (\textcolor{Myorange}{primary} or \textcolor{Myblue}{secondary}).


\subsection{Do Captions Help at All?}\label{captions}

\edit{In Borkin et al.\cite{Borkin2016BeyondMV}, participants viewed different visualizations for 10 seconds each while their eyes were tracked. After that, they looked at the same visualizations blurred by a Gaussian filter and recalled what they memorized from the visualizations. Borkin et al.\cite{Borkin2016BeyondMV} found that textual elements, including titles and captions, were mentioned the most in participants' responses and were key elements in helping participants memorize visualizations.}
 
\begin{figure}[t]
\includegraphics[width=8.5cm]{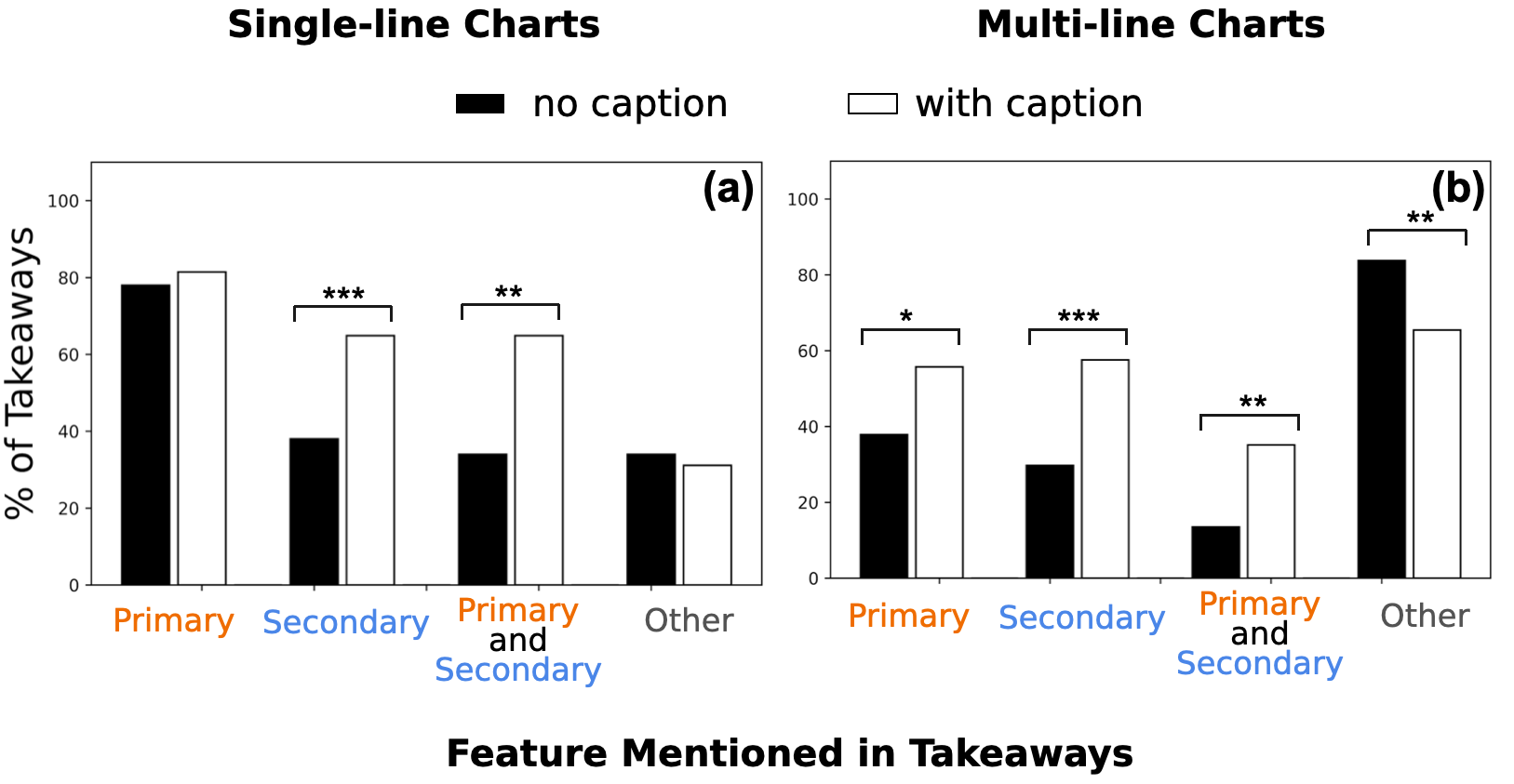}
\caption{\edit{Percentage of takeaways that mention a feature for chart types with and without captions. *, **, and *** denote a p-value of ${<}0.05$, ${<}0.01$, and ${<}0.001$ from Welch's t-test after Bonferonni correction.}  
}
\label{f:result0}
\end{figure}

\edit{However, Borkin et al. \cite{Borkin2016BeyondMV} did not look into the effect of textual elements on helping participants memorize specific features of visualizations.  Therefore, as a first analysis, we examined whether adding a caption can affect the feature mentioned in participants' takeaways. \Cref{f:result0} reports the percentage of takeaways that mention a feature for single- and multi-line charts with and without captions. We aggregated the data for all 4 caption types to calculate the results for the with caption group. Due to the disparity in sample sizes (no caption group account for 20\%, while with caption group account for 80\%), we used Welch's t-test, which doesn't assume equal variance, to check whether there is a significant difference within each pair. \Cref{f:result0} shows significant differences after Bonferonni correction. }

\edit{The result shows that adding a caption significantly increases the percentage of takeaways that mention secondary features in single-line charts and both features in multi-line charts. However, it is not the case for the primary features of single-line charts. A possible reason is that the primary features of single-line charts are already very salient without a caption; hence, the improvement from adding a caption is marginal\cite{Kim2021TowardsUH}. }

Additionally, many participants mentioned other features in their takeaways for multi-line charts when there was no caption. A potential explanation is that multi-line charts are more complex and contain more features. Significantly fewer participants did so when there was a caption, suggesting that participants focused more on the captioned (\textcolor{Myorange}{primary} or \textcolor{Myblue}{secondary}) features.



\subsection{Semantic Level and Feature Salience}

We now focus on conditions where there is a caption, and study how the feature and semantic level of the caption affect memorability.   Specifically, we hypothesized:
\edit{
\begin{myitemize}
    \item \textbf{H1.1:} for \textbf{single-line} charts, semantic level and captioned feature affect if the takeaway mentions the \textcolor{Myorange}{\textbf{primary}} feature.
    \item \textbf{H1.2:} for \textbf{multi-line} charts, semantic level and captioned feature affect if the takeaway mentions the \textcolor{Myorange}{\textbf{primary}} feature.
    \item \textbf{H1.3:} for \textbf{single-line} charts, semantic level and captioned feature affect if the takeaway mentions the \textcolor{Myblue}{\textbf{secondary}} feature.
    \item \textbf{H1.4:} for \textbf{multi-line} charts, semantic level and captioned feature affect if the takeaway mentions the \textcolor{Myblue}{\textbf{secondary}} feature.
\end{myitemize}
}

To this end, we partitioned the responses by chart type and feature mentioned in the takeaway. \edit{An F-test reported equal variances across partitions.} For each partition, we ran a two-way ANOVA test to analyze the effect of semantic level and captioned feature on whether the takeaway mentions a primary/secondary feature.  

\stitle{ANOVA Results:}
\edit{
For single-line charts, the tests revealed that semantic level and feature each have a statistically significant effect on whether the takeaway mentions the primary feature, but only the caption feature has a significant effect on whether the takeaway mentioning the secondary feature.
For multi-line charts, the tests revealed that semantic level and feature each have a statistically significant effect on whether the takeaway mentions the primary or secondary feature.    These findings were corroborated by a post-hoc Tukey HSD test.    
\Cref{t:anova} summarizes the results: the \green{first column} shows the dependent variables (each chart type and takeaway feature), \green{the second column} shows the independent variables \green{(captioned feature, semantic level, and their interaction)}, and the last column shows the corresponding p-values.
}

\begin{table}[t]
\centering
\small
\begin{tabular}{m{0.15\columnwidth} > {\centering}m{0.12\columnwidth} > 
{\centering}m{0.03\columnwidth} > 
{\centering}m{0.07\columnwidth} > 
{\centering}m{0.07\columnwidth}>{\centering}m{0.08\columnwidth}>{\centering\arraybackslash}m{0.15\columnwidth}}
\textbf{Feature in {Takeaway}}  & \textbf{Sources of Variation} & \textbf{Df} & \textbf{Sum Sq} & \textbf{Mean Sq} & \textbf{F} & \textbf{Pr($>$F)} \\
\midrule
\multirow{3}{1.2cm}{\textbf{\textcolor{Myorange}{Primary}} 
in \textbf{Single-line Charts}} & Captioned Feature & $1$ & $1.231$  & $1.231$  & $8.667$ & $\mathbf{0.004^{**}}$\\ 
\cmidrule{2-7} & Semantic Level & $1$ &  $0.665$  & $0.665$ &   $4.686$ & $\mathbf{0.032^{*}}$ \\\cmidrule{2-7} & Interaction & $1$ &  $0.040$  & $0.040$ &   $0.278$ & $0.599$\\
\midrule
\multirow{3}{1.4cm}{\textbf{\textcolor{Myblue}{Secondary}} in \textbf{Single-line Charts}} & Captioned Feature & $1$ &  $1.020$ & $1.023$ &  $4.580$ & $\mathbf{0.034^{*}}$\\ 
\cmidrule{2-7} & Semantic Level & $1$ &   $0.410$ &  $0.413$ &  $1.848$ &  $0.176$ \\\cmidrule{2-7} & Interaction & $1$ &  $0.120$ &  $0.118$ &   $0.527$ &  $0.469$\\
\midrule
\multirow{3}{1.2cm}{\textbf{\textcolor{Myorange}{Primary}} in \textbf{Multi-line Charts}} & Captioned Feature & $1$ &  $5.030$ &   $5.027$ &  $23.339$ & $\mathbf{<0.001^{***}}$\\ 
\cmidrule{2-7} & Semantic Level & $1$ &  $0.980$ &   $0.981$ &   $4.553$ &  $\mathbf{0.0344^{*}}$ \\\cmidrule{2-7} & Interaction & $1$ &  $0.010$ &   $0.015$ &   $0.067$ &   $0.795$\\
\midrule
\multirow{3}{1.4cm}{\textbf{\textcolor{Myblue}{Secondary}} in \textbf{Multi-line Charts}} & Captioned Feature & $1$ &  $2.970$ &  $2.969$ &  $13.452$ & $\mathbf{<0.001^{***}}$\\ 
\cmidrule{2-7} & Semantic Level & $1$ &   $1.790$ &  $1.793$ & $8.124$ & $\mathbf{0.005^{**}} $ \\\cmidrule{2-7} & Interaction & $1$ &  $0.010$ & $0.014$ & $0.062$ & $0.803$\\
\midrule
\end{tabular}
\vspace{1mm}
\caption{ANOVA test results for feature mentioned in takeaways (first column) and different sources of variations (second column). Tests for semantic level of captions, feature described in captions, and their interactions are performed without the no caption control group. *, **, and *** denote a p-value of ${<}0.05$, ${<}0.01$, and ${<}0.001$. }
\label{t:anova}
\end{table}

\begin{figure}[t]
\includegraphics[width=8.5cm]{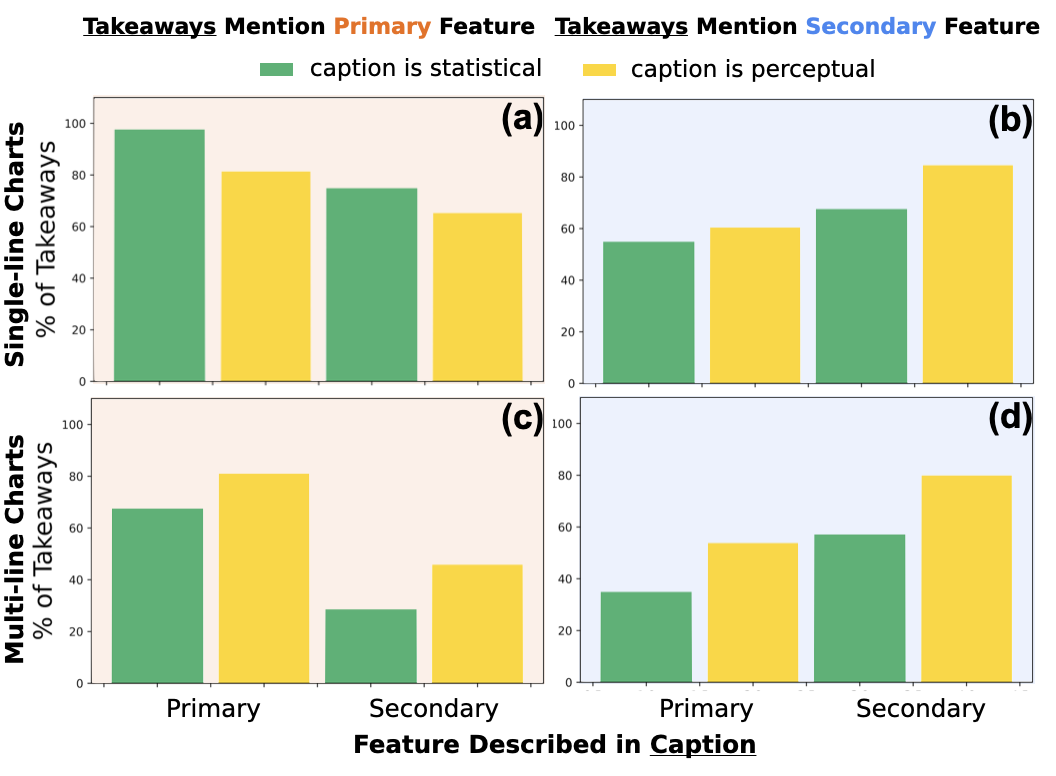}
\caption{Percentage of takeaways that mention a feature for caption types (x-axis). \edit{100\% of each bar is $40\pm{5}$ takeaways.}
}
\label{f:result1}
\end{figure}

\edit{\Cref{f:result1}  also reports the percentage of
\green{takeaways} that mention the primary/secondary feature (columns) for single- and multi-line charts (rows), as we vary the caption type (x-axis). The bars are colored with respect to the semantic levels of the captions, and the background of each subfigure is lightly colored to indicate the feature (\textcolor{Myorange}{primary} or  \textcolor{Myblue}{secondary}) mentioned in the takeaways.} 

\stitle{Overall Patterns:}
\green{\Cref{f:result1} shows that memorability is generally lower in multi-line (\Cref{f:result1}(c,d)) than single-line charts (\Cref{f:result1}(a,b)), potentially because they are more complex and no single feature ``stands out''~\cite{Carpenter1998AMO}. } 
\green{\Cref{f:result1}(a) shows that when we consider single-line charts, the user is more likely to mention the primary feature in their takeaway when the caption is statistical rather than perceptual (the first two bars).  This pattern holds irrespective of the feature described in the caption (the first two bars, the last two bars).   In contrast, for all other conditions (\Cref{f:result1}(b,c,d)) perceptual phrasing results in higher memorability than statistical phrasing.   From \Cref{t:anova}, these differences between semantic levels are  significant in \Cref{f:result1}(a,c,d), and not significant in \Cref{f:result1}(b).}


\stitle{Feature Salience:} 
\edit{Kim et al. \cite{Kim2021TowardsUH} examined the features mentioned in participants' takeaways when a caption mentions first, second, third and non-prominent features of single-line charts respectively. Their results showed that if a caption mentions first and second prominent features, people are more likely to mention the feature in caption as their takeaway.}

\edit{While Kim et al.\cite{Kim2021TowardsUH} only studied single-line charts, we expanded the chart type to include both single- and multi-line charts. Our results for single-line charts are consistent with Kim et al.'s \cite{Kim2021TowardsUH}. For multi-line charts, our results also show that for both primary and secondary features, people are significantly more likely to mention the captioned feature as their takeaway.}

\stitle{Semantic Levels in Single-line Charts: }
\edit{In \Cref{captions}, we found that adding a caption does not affect whether participants mention the primary features of single-line charts.  However,  since we did not differentiate whether the caption explicitly mentioned the primary or secondary feature in \Cref{captions}, a given feature can be less memorable simply because it was not mentioned in the caption (the last two bars in \Cref{f:result1}(a)), as also observed by Kim et al.~\cite{Kim2021TowardsUH}. This effect might have made the overall effect of adding a caption insignificant in \Cref{captions}.}

\edit{To build on this, the ANOVA test (\Cref{t:anova}, block 1) shows that, when the visualization includes a caption, the semantic level has an effect in increasing {\it the primary feature} mentioned for {\it single-line} charts. Specifically, in single-line charts, the primary feature is significantly more memorable when described at the statistical level than at the perceptual level. The effect of the semantic level persists irrespective of the feature described in the caption.}

\green{\Cref{captions} also shows that adding a caption increases the memorability of the secondary feature for single-line charts. Nonetheless, the ANOVA test (\Cref{t:anova}, block 2) shows that caption's semantic level has no significant effect on the takeaway features. This result suggests that in single-line charts, the secondary feature is significantly more memorable as long as the caption describes it, regardless of the caption's semantic level.}

\stitle{Semantic Levels in Multi-line Charts:}
\green{What happens when the charts are more complex? For both primary and secondary features of multi-line charts, \Cref{captions} shows that the presence of a caption can significantly increase their memorability. The ANOVA results (\Cref{t:anova}, block 3 and 4) suggest that both captioned feature and semantic level play a role. Captions at perceptual level improve participants' memorability of both primary and secondary features when they are mentioned in the captions.}

\stitle{Assessing Hypotheses: }
\edit{Considering the results for feature salience and semantic levels, we can accept \textbf{H1.1},  \textbf{H1.2} and \textbf{H1.4}. However, as ANOVA test does not show significance for the effect of semantic level in takeaway features (\Cref{t:anova}, block 2), we cannot accept \textbf{H1.3} even though the captioned feature affects if a takeaway mentions the secondary feature in single-line charts. A possible reason is that single-line charts are less complex than multi-line charts, and thus their secondary feature can be recognized and recalled as long as a caption that describes it is present.}

\subsubsection{Reliance on Captions}
\begin{figure}[t]
\centering
\includegraphics[width=8.5cm]{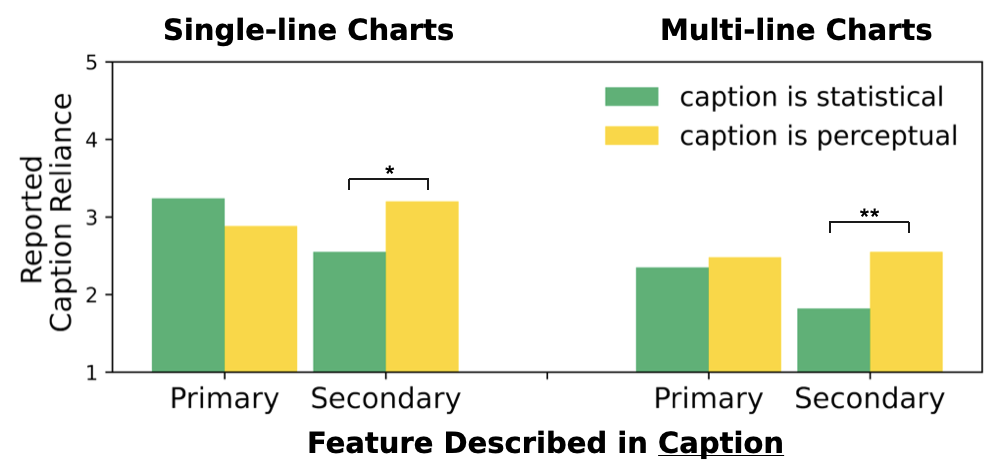}
\caption{Self-reported reliance on different caption types for single- and multi-line charts. 1 means not dependent and 5 means entirely dependent. *, ** denote a p-value of $<0.05$ and $<0.01$, respectively.}
\label{f:reliance}
\end{figure}

Along with takeaways, participants also reported their reliance on the chart and caption \edit{on a 5-Likert scale}.  \Cref{f:reliance}  shows the mean of caption reliance for each chart-caption pair. As the reliance data is ordinal, we ran a Mann-Whitney U test to compare the rank sum of reliance on statistical and perceptual captions when they describe the same feature for each chart type. \Cref{f:reliance} shows that when the caption describes the primary feature, participants relied on the caption the same amount irrespective of the semantic level.  In contrast, when it describes the secondary feature, participants relied more on perceptual captions (single-line: Mann-Whitney $U=626, p=0.042$; multi-line: Mann-Whitney $U=1116, p=0.008$; \green{a greater U indicates a greater difference in rank sums between self-reported reliance on statistical and perceptual captions in each group). }

This result reflects participants' subjective evaluation of the effectiveness of different caption types. While the ANOVA test in \Cref{captions} does not show significant difference between statistical and perceptual level captions for secondary features of single-line charts, the significance shown by reliance difference suggests that participants might consider a perceptual level caption more effective.


\subsection{What Types of Numbers do Participants Remember?}

\edit{We then performed an exploratory analysis to understand the types of numbers that participants remembered.  Specifically, since the points are two dimensional, we computed whether a takeaway referenced the x-axis (year) or y-axis (measure) value of a point in the chart. We formulated these as two hypotheses:}

\begin{myitemize}
\item \edit{\textbf{H1.1:} including a caption has an affect on whether the user's takeaway includes an x-axis value}
\item \edit{\textbf{H1.2:} including a caption has an affect on whether the user's takeaway includes a y-axis value.}
\end{myitemize}

\edit{We evaluate these hypotheses using fisher's exact test.  We find that the use of a caption does not affect the likelihood of referencing x-axis values in the takeaways ($p=0.5316$), but does affect that for y-axis values ($p=0.0238$). For these reasons, we do not find evidence to support \textbf{H1.1} and accept \textbf{H1.2}.}

\begin{figure}[h]
\centering
\includegraphics[width=8.5cm]{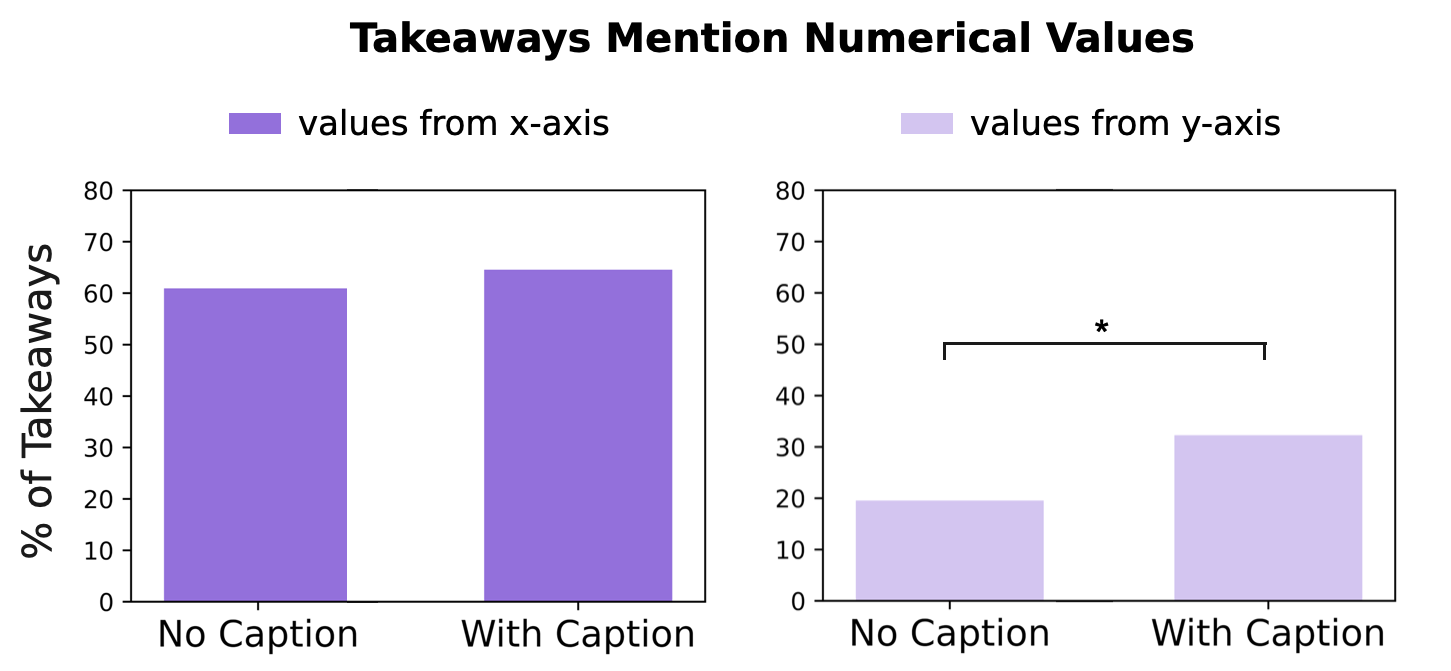}
\caption{Percentage of takeaways that mention an x- or y-axis value of a point. * indicates statistical significance ($p<0.05$) over the corresponding no-caption bar.}
\label{fig:numinfo}
\end{figure}

\edit{\Cref{fig:numinfo} plots percentage of takeaways containing x/y-axis values under each condition.   We can see that 
even without a caption, participants were far more likely to mention x-axis rather than the y-axis values. A potential reason is that captions (and possibly participants) naturally refer to the x-axis.  For instance, trend descriptions such as ``increase between 2000 and 2010'' naturally refer to x-axis values.  We further ran a fisher exact test among the four caption types (semantic levels $\times$ captioned feature) and did not find a significant effect ($p=0.3495$ for x-axis, $p=0.96$ for y-axis in takeaways).}

\section{\edit{Conclusion}}\label{s:conclusion}
We examine how captions \edit{of different features and} varying semantic levels can affect the features mentioned in the readers' takeaways. These results lead to recommendations for composing captions, depending on what features the author wishes to emphasize and convey.   

We find that readers tend naturally to remember a highly salient feature in a chart, such as an extremum or abrupt change in a single-line chart.   Yet, the author can further emphasize the primary feature by focusing the caption on the statistical properties of the feature.  Meanwhile, captions can also focus the reader on medium or low-salience features---these may be secondary trends or any feature in a complex multi-line chart.  \edit{For these lower salience features, and particularly in multi-line charts,} perceptual level captions increase memorability of the described feature.     To summarize these recommendations:

\begin{myitemize}
\item \textbf{Use Statistical Level} for high salience features in single-line charts. 

\item \textbf{Use Perceptual Level} for medium/low salience features in multi-line charts.
\end{myitemize}



\stitle{Limitations:} \edit{First, the participants vary in visualization reading expertise, and can have different interpretations of instructions for the study. Although we tried to minimize the implication by having a screening test and manually filtering out responses that clearly misinterpret the study, this factor can still introduce noises in our study. Second, a larger pool of participants is needed to explicate some findings in this study, including the effect of various caption types on helping recall numerical information.}

\stitle{Applications:} Our findings can potentially help data visualization practitioners write captions to more effectively communicate patterns that they want to highlight. They can also help standardize guidelines for machine-generated captions. Although this study only involves sighted users, we hope this approach of categorizing and analyzing captions' semantic levels can be applied to exploring effective alt-text writing for the visually impaired community as well.


\bibliographystyle{abbrv-doi-hyperref-narrow}

\bibliography{template}

\begin{thebibliography}{10}
\renewcommand*{\sfdefault}{PTSansNarrow-TLF}

\bibitem{Borkin2016BeyondMV}
\href{https://doi.org/10.1109/TVCG.2015.2467732}{M.~A. Borkin, Z.~Bylinskii,
  N.~W. Kim, C.~M. Bainbridge, C.~S. Yeh, D.~Borkin, H.~Pfister, and A.~Oliva}.
\newblock \href{https://doi.org/10.1109/TVCG.2015.2467732}{Beyond memorability:
  Visualization recognition and recall}.
\newblock \href{https://doi.org/10.1109/TVCG.2015.2467732}{{\em IEEE
  Transactions on Visualization and Computer Graphics}},
  \href{https://doi.org/10.1109/TVCG.2015.2467732}{22(1):519--528},
  \href{https://doi.org/10.1109/TVCG.2015.2467732}{2016}.
  \href{https://doi.org/10.1109/TVCG.2015.2467732}
{doi: \textsf{%
10\hspace{.1pt}\discretionary{.}{%
}{.}\hspace{.4pt}1109\discretionary{/}{%
}{/}TVCG\hspace{.1pt}\discretionary{.}{%
}{.}\hspace{.4pt}2015\hspace{.1pt}\discretionary{.}{%
}{.}\hspace{.4pt}2467732}}


\bibitem{Carpenter1998AMO}
\href{https://doi.org/10.1037/1076-898X.4.2.75}{P.~A. Carpenter and P.~Shah}.
\newblock \href{https://doi.org/10.1037/1076-898X.4.2.75}{A model of the
  perceptual and conceptual processes in graph comprehension}.
\newblock \href{https://doi.org/10.1037/1076-898X.4.2.75}{{\em Journal of
  Experimental Psychology: Applied}},
  \href{https://doi.org/10.1037/1076-898X.4.2.75}{4(2):75--100},
  \href{https://doi.org/10.1037/1076-898X.4.2.75}{1998}.
  \href{https://doi.org/10.1037/1076-898X.4.2.75}
{doi: \textsf{%
10\hspace{.1pt}\discretionary{.}{%
}{.}\hspace{.4pt}1037\discretionary{/}{%
}{/}1076\discretionary{%
}{-}{-}898X\hspace{.1pt}\discretionary{.}{%
}{.}\hspace{.4pt}4\hspace{.1pt}\discretionary{.}{%
}{.}\hspace{.4pt}2\hspace{.1pt}\discretionary{.}{%
}{.}\hspace{.4pt}75}}


\bibitem{diagramcenter}
D.~Center.
\newblock Specific guidelines - graphs, 2019.
\newblock \url{http://diagramcenter.org/specific-guidelines-e.html}.

\bibitem{CFPB}
{CFPB Design System}.
\newblock Data visualization guidelines, 2022.
\newblock
  \url{https://cfpb.github.io/design-system/guidelines/data-visualization-guidelines}.

\bibitem{Jung2021CommunicatingVW}
\href{https://doi.org/10.1109/TVCG.2021.3114846}{C.~Jung, S.~Mehta,
  A.~Kulkarni, Y.~Zhao, and Y.-S. Kim}.
\newblock \href{https://doi.org/10.1109/TVCG.2021.3114846}{Communicating
  visualizations without visuals: Investigation of visualization alternative
  text for people with visual impairments}.
\newblock \href{https://doi.org/10.1109/TVCG.2021.3114846}{{\em IEEE
  Transactions on Visualization and Computer Graphics}},
  \href{https://doi.org/10.1109/TVCG.2021.3114846}{28(1):1095--1105},
  \href{https://doi.org/10.1109/TVCG.2021.3114846}{2022}.
  \href{https://doi.org/10.1109/TVCG.2021.3114846}
{doi: \textsf{%
10\hspace{.1pt}\discretionary{.}{%
}{.}\hspace{.4pt}1109\discretionary{/}{%
}{/}TVCG\hspace{.1pt}\discretionary{.}{%
}{.}\hspace{.4pt}2021\hspace{.1pt}\discretionary{.}{%
}{.}\hspace{.4pt}3114846}}


\bibitem{Kildal2006NonvisualOO}
\href{https://doi.org/10.1145/1125451.1125634}{J.~Kildal and S.~A. Brewster}.
\newblock \href{https://doi.org/10.1145/1125451.1125634}{Non-visual overviews
  of complex data sets}.
\newblock \href{https://doi.org/10.1145/1125451.1125634}{{\em CHI '06 Extended
  Abstracts on Human Factors in Computing Systems}},
  \href{https://doi.org/10.1145/1125451.1125634}{p. 947–952},
  \href{https://doi.org/10.1145/1125451.1125634}{2006}.
  \href{https://doi.org/10.1145/1125451.1125634}
{doi: \textsf{%
10\hspace{.1pt}\discretionary{.}{%
}{.}\hspace{.4pt}1145\discretionary{/}{%
}{/}1125451\hspace{.1pt}\discretionary{.}{%
}{.}\hspace{.4pt}1125634}}


\bibitem{Kim2021TowardsUH}
\href{https://doi.org/10.1145/3411764.3445443}{D.~H. Kim, V.~Setlur, and
  M.~Agrawala}.
\newblock \href{https://doi.org/10.1145/3411764.3445443}{Towards understanding
  how readers integrate charts and captions: A case study with line charts}.
\newblock \href{https://doi.org/10.1145/3411764.3445443}{{\em Proceedings of
  the 2021 CHI Conference on Human Factors in Computing Systems}},
  \href{https://doi.org/10.1145/3411764.3445443}{2021}.
  \href{https://doi.org/10.1145/3411764.3445443}
{doi: \textsf{%
10\hspace{.1pt}\discretionary{.}{%
}{.}\hspace{.4pt}1145\discretionary{/}{%
}{/}3411764\hspace{.1pt}\discretionary{.}{%
}{.}\hspace{.4pt}3445443}}


\bibitem{Kim2021AccessibleVD}
\href{https://doi.org/10.1111/cgf.14298}{N.~W. Kim, S.~C. Joyner,
  A.~Riegelhuth, and Y.-S. Kim}.
\newblock \href{https://doi.org/10.1111/cgf.14298}{Accessible visualization:
  Design space, opportunities, and challenges}.
\newblock \href{https://doi.org/10.1111/cgf.14298}{{\em Computer Graphics
  Forum}}, \href{https://doi.org/10.1111/cgf.14298}{40(3):173--188},
  \href{https://doi.org/10.1111/cgf.14298}{2021}.
  \href{https://doi.org/10.1111/cgf.14298}
{doi: \textsf{%
10\hspace{.1pt}\discretionary{.}{%
}{.}\hspace{.4pt}1111\discretionary{/}{%
}{/}cgf\hspace{.1pt}\discretionary{.}{%
}{.}\hspace{.4pt}14298}}


\bibitem{Kong2019TrustAR}
\href{https://doi.org/10.1145/3290605.3300576}{H.~K. Kong, Z.~Liu, and
  K.~Karahalios}.
\newblock \href{https://doi.org/10.1145/3290605.3300576}{Trust and recall of
  information across varying degrees of title-visualization misalignment}.
\newblock \href{https://doi.org/10.1145/3290605.3300576}{{\em Proceedings of
  the 2019 CHI Conference on Human Factors in Computing Systems}},
  \href{https://doi.org/10.1145/3290605.3300576}{2019}.
  \href{https://doi.org/10.1145/3290605.3300576}
{doi: \textsf{%
10\hspace{.1pt}\discretionary{.}{%
}{.}\hspace{.4pt}1145\discretionary{/}{%
}{/}3290605\hspace{.1pt}\discretionary{.}{%
}{.}\hspace{.4pt}3300576}}


\bibitem{mitdataset}
A.~Lundgard and A.~Satyanarayan.
\newblock Visualization text model open source data.
\newblock \url{http://vis.csail.mit.edu/pubs/vis-text-model/data/}.

\bibitem{2022-vis-text-model}
\href{https://doi.org/10.1109/TVCG.2021.3114770}{A.~Lundgard and
  A.~Satyanarayan}.
\newblock \href{https://doi.org/10.1109/TVCG.2021.3114770}{{Accessible
  Visualization via Natural Language Descriptions: A Four-Level Model of
  Semantic Content}}.
\newblock \href{https://doi.org/10.1109/TVCG.2021.3114770}{{\em IEEE
  Transactions on Visualization \& Computer Graphics (Proceedings of IEEE
  VIS)}}, \href{https://doi.org/10.1109/TVCG.2021.3114770}{2022}.
  \href{https://doi.org/10.1109/TVCG.2021.3114770}
{doi: \textsf{%
10\hspace{.1pt}\discretionary{.}{%
}{.}\hspace{.4pt}1109\discretionary{/}{%
}{/}TVCG\hspace{.1pt}\discretionary{.}{%
}{.}\hspace{.4pt}2021\hspace{.1pt}\discretionary{.}{%
}{.}\hspace{.4pt}3114770}}


\bibitem{Moraes2014EvaluatingTA}
\href{https://doi.org/10.1145/2661334.2661368}{P.~S. Moraes, G.~Sina, K.~F.
  McCoy, and S.~Carberry}.
\newblock \href{https://doi.org/10.1145/2661334.2661368}{Evaluating the
  accessibility of line graphs through textual summaries for visually impaired
  users}.
\newblock \href{https://doi.org/10.1145/2661334.2661368}{{\em Proceedings of
  the 16th international ACM SIGACCESS conference on Computers \&
  accessibility}}, \href{https://doi.org/10.1145/2661334.2661368}{2014}.
  \href{https://doi.org/10.1145/2661334.2661368}
{doi: \textsf{%
10\hspace{.1pt}\discretionary{.}{%
}{.}\hspace{.4pt}1145\discretionary{/}{%
}{/}2661334\hspace{.1pt}\discretionary{.}{%
}{.}\hspace{.4pt}2661368}}


\bibitem{ryan2018glance}
G.~Ryan, A.~Mosca, R.~Chang, and E.~Wu.
\newblock At a glance: Pixel approximate entropy as a measure of line chart
  complexity.
\newblock {\em IEEE transactions on visualization and computer graphics},
  25(1):872--881, 2018.

\bibitem{W3C}
W3C.
\newblock Complex images, 2019.
\newblock \url{https://www.w3.org/WAI/tutorials/images/complex/}.

\bibitem{GBH}
{WGBH National Center for Accessible Media}.
\newblock Effective practices for description of science content within digital
  talking books, 2008.
\newblock
  https://www.wgbh.org/foundation/ncam/guidelines/effective-practices-for-description-of-science-content-within-digital-talking-books.

\end{thebibliography}
\end{document}